\documentclass[12pt,preprint]{aastex}

\begin{document}
\title{Detection of Galaxy Spin Alignments in the PSCz Shear Field}
\author{Jounghun Lee}
\affil{Institute of Astronomy and Astrophysics, Academia Sinica,
Taipei, Taiwan}
\email{taiji@asiaa.sinica.edu.tw}
\author{Ue-Li Pen}
\affil{Canadian Institute for Theoretical Astrophysics,
Toronto, Ont. M5S 3H8, Canada}
\email{pen@cita.utoronto.ca}

\begin{abstract} 

We report the first  direct observational evidence for alignment
of galaxy spin axes with the local tidal shear field.   We measure
quantitatively the strength of this directional correlation of disk
galaxies from the Tully catalog with the local shear field reconstructed
from PSCz galaxies.  We demonstrate that the null hypothesis of random
galaxy alignments relative to the shear frame can be ruled out more than
$99.98\%$ confidence. The observed intrinsic correlation averaged over
the censored samples that have detected non-zero signals is measured in 
terms of the alignment parameter $\bar{a}=0.17\pm 0.04$, which includes 
only statistical errors of the censored data, but not the cosmic variance 
error. The reconstruction procedure is likely to underestimates $\bar{a}$ 
systematically.  
Our result is consistent with the linear tidal torque model, 
and supports the idea that the present galaxy spins may be used 
as a probe of primordial tidal shear and mass
density fields.  The intrinsic alignments of galaxy spins dominate over 
weak gravitational lensing for shallow surveys such like SDSS, while 
it should be negligible for deeper surveys at $z \sim 1$. 

\end{abstract} 

\newcommand{\etal}{{\it et al.}}
\newcommand{\beq}{\begin{equation}}
\newcommand{\eeq}{\end{equation}}
\newcommand{\ben}{\begin{eqnarray}}
\newcommand{\een}{\end{eqnarray}}
\newcommand{\bL}{{\bf L}}
\newcommand{\bk}{{\bf k}}
\newcommand{\bx}{{\bf x}}
\newcommand{\bxp}{{\bf x}^{\prime}}
\newcommand{\bI}{{\bf I}}
\newcommand{\bT}{{\bf T}}
\newcommand{\del}{\delta}
\newcommand{\sig}{\sigma}
\newcommand{\hlam}{\hat{\lambda}}
\newcommand{\hL}{\hat{L}}

\section{INTRODUCTION}

In the standard gravitational instability paradigm, the linear tidal
torque theory explains the origin of the angular momentum of a galaxy
as arising from its tidal coupling with the surrounding matter at the
early proto-galactic stage \citep{hoy49,pee69,dor70,whi84}, and predicts
a correlation of its spin axis with the intermediate principal axis of
local shear \citep{dub92,cat-the96,lp00}.  In this model, each component
of the proto-galaxy angular momentum vector in the shear principal axis
frame is represented by a difference of a pair of the three eigenvalues
of the shear tensor, and thus the intermediate component proportional
to the difference between the largest and the smallest eigenvalues
of the shear tensor dominates the direction of the angular momentum.
This correlation of proto-galaxy spin axis with the initial tidal shear
field is intrinsic and local, and distinct from the weak gravitational
lensing effect \cite{cri-etal01}.

An important question is whether the intrinsic alignments of
proto-galaxy spins with the tidal shear fields remain till present
or not. If so, it will have a considerable impact on weak lensing
searches since the intrinsic shear effect on galaxy spins will
play a role of systematic errors.  Furthermore recently Lee \& Pen
(2000,2001) have shown that the relatively easily measurable galaxy
spin field can be used to reconstruct the initial tidal shear and mass
density fields provided that the effect of intrinsic shear on galaxy
spins persist.  
Several numerical tests \citep{hea-etal00,por-etal01a,por-etal01b} and 
observational searches \citep{pen-etal00,bro-etal00}
have so far found some evidences for the existence of intrinsic spin 
alignments, but early attempts were all indirect.

The detection of intrinsic correlations between the tidal shear and galaxy 
spin fields from the observed universe requires independent measurement of 
the two fields. Here we attempt to measure the tidal shear and galaxy spin 
fields from two independent all-sky catalogs and measure the local  
directional correlations of the two fields directly. 

\section{FIELD MEASUREMENTS} 

One would like to measure the intrinsic shear in the initial Lagrangian
field, which is, however, practically impossible to derive from present
observational data.  What one can do at most is to measure the evolved 
final Eulerian shear and approximate the Lagrangian shear to
it. This approximation can be improved by applying a Wiener filter to
the observational data.

For the measurement of shear, we have used the PSCz (a complete redshift 
survey from the IRAS Point Source Catalog) data, which provides a flux-limited
all-sky catalog of 15500 galaxies detected by IRAS. 
It covers $84 \%$ of the sky down to a flux limit of $0.6$ Jy at $60 \mu$m.  
The remaining $16 \%$ of the sky is a gap region containing the galactic 
plane and survey gaps excluded in the PSCz \cite{sau-etal00}.  

The flux-limited catalog is not a homogeneous sample since the fraction
of galaxies observed decreases with distance. For the measurement
of tidal shear, it is statistically simpler to use a volume-limited
sample where the fraction of galaxies is constant with distance.
We converted the flux limited PSCz catalog into volume-limited samples by
dividing the whole volume into the 7 spherical shells of distance
range $[0,50],[50,75],[75,100],[100,125]$,$[125,150]$,\\
$[150,175]$ and $[175,200]$ (in unit  of $h^{-1}$Mpc) as in the previous 
literature \citep{yah-etal91,sza-etal00} 
and including only galaxies luminous enough to be above 
the flux limit if placed at the outer edge of each volume. We
ended up with 7 volume-limited samples containing $1730,1541,1356$,
$1001,701,586$ and $442$ galaxies respectively.

We first measure the raw density field $\delta(\bx)$ of a chosen PSCz sample 
(say, $V$), assuming that the mass density is proportional to the number 
density of galaxies:  
$\del(\bx) = \frac{1}{\rho_{0}}\sum_{\bx_{i} \in V} 
\del_{D}(\bx -\bx_{i}) - 1$, 
where $\delta_{D}$ is the Dirac-delta function, and $\rho_{0}$ is the mean 
number density of galaxies of the sample $V$.  We calculate the raw density 
field of each sample separately as if there were no mass source outside the 
sample under consideration. 

To minimize the shot noise due to the finite number of mass sources and
optimize the approximation of the Lagrangian shear to the Eulerian shear,
we convolve this density field with the Wiener filter, based on the
standard power law correlation function for the galaxies.  We find the
Fourier space Wiener filter to be $W(k) = \frac{Ak^{-1.2}}{Ak^{-1.2} +
\rho_{0}^{-1}}$ with $A = 0.903\times 2\pi^{2}\times(4.43)^{1.8}$. Here
$\rho_{0}^{-1}$ is the noise power spectrum derived from the Poissonian
distribution.  Note that each sample has a different Wiener filter
distinguished by a different value of $\rho_{0}$.

\newcommand{\tW}{\tilde{W}}
Since the Wiener filtered gravitational potential in Fourier space is 
$\phi(\bk) = k^{-2}\delta(\bk)W(k)$,  its real counter-part is written 
as $\phi(\bx) = \int_{V}\delta(\bxp)\tW(\bx - \bxp) d^{3}\bxp$ where
$\tW$ is a modified Wiener filter defined in Fourier space 
as $\tW(k) = k^{-2}W(k)$. 

Now we find the gravitational potential due to the mass 
density of the chosen sample $V$ at some field point $\bx$:   
\ben 
\phi(\bx) &=& \frac{1}{2\pi^{2}\rho_{0}}\sum_{\bx_{i} \in V}\int
k \tW(k)\frac{\sin(k|\bx - \bx_{i}|)}{|\bx - \bx_{i}|}dk  , 
\nonumber \\
&&+\frac{2}{\pi}\int \tW(k)
\bigg{\{}\frac{R_1}{k}\cos(R_{1}k) - \frac{R_2}{k}\cos(R_{2}k) + 
\frac{1}{k^2}\sin(R_{2}k) - \frac{1}{k^2}\sin(R_{1}k)\bigg{\}}dk,
\label{eqn:pdet}
\een 
where $R_1$ and $R_2$ are the inner and outer radii of $V$.  The total 
gravitational potential at $\bx$ will be the super-positions of all the 
gravitational potentials calculated from the seven PSCz samples separately.  
The tidal shear field can be finally determined as the second derivative 
of the total gravitational potential such that  
$T_{ij}(\bx)=\sum_{\rm{samples}}\partial_{i}\partial_{j}\phi(\bx)$.  

There are, however, two sources of bias in the measurement of tidal 
shear: the existence of a gap region excluded in the PSCz and the 
redshift distortion effect on the orientation of the shear principal 
axes along the line of sight. 
To minimize these errors, first we compensate for the existence 
of the gap by adding the gap contribution to the total mass density, 
assuming a uniform mass distribution in the gap. Second, we cancel out 
the redshift distortion bias in the determination of shear by readjusting 
the shear principal axes to have a statistically isotropic distribution. 
Basically we force the angle between the shear principal 
axes of each shear and the line-of-sight direction to have a uniform 
distribution in the range of $[0,2\pi]$ without changing its original 
ranking order.  The most important intermediate axis has been adjusted 
first, the major axis as next, and finally the minor axis has been 
automatically determined from the orthogonality condition.  

Note that these procedures minimize the systematic bias in the
measurement of shear orientation but also lower the inferred
strength of intrinsic shear effects since the shear axes may
deviate further from the true orientations.
Finally we renormalize the resulting shears into the unit traceless shear
since our eventual interest is not in the magnitude but the axis 
orientation of the tidal shear. 

Regarding the galaxy spin field, we consider the unit spins of 12122 
Tully disk galaxies as used by Pen \etal\ (2001). Let us  
briefly review the work of Pen \etal .
The unit spin vector of each disk galaxy is initially estimated from 
the major-minor axis ratio and the position angle,  
assuming that a disk galaxy is an infinitely thin disk with its spin axis 
perpendicular to the disk plane. 
The measured axis ratio and position angle of a disk galaxy determine the 
inclination and orientation angles of the disk.
The inclination angle is only determined up to a sign ambiguity.  
We take into account this sign ambiguity in the determination 
of line-of-sight component of a unit spin,  and construct two unit spin 
vectors at each Tully galaxy position. Therefore we end up with twice as 
many unit spins as Tully galaxies. 

Real disk galaxies, however, are not infinitely thin disks. Their
intrinsic thickness and deviation in shape from a perfect ellipse create
biases in the statistical distribution of inclination angles.
Although this statistical bias has a different source from the redshift 
distortions, its effect on the galaxy spin axis can be removed by the 
same procedure that we use for the adjustment of shear principal axis.  
We adjust $\alpha$ to have a uniform distribution in range 
of $[0,\pi]$, as would be the unbiased case. 

This statistical bias might depend on both the galaxy
distance and Hubble type (intrinsic morphology).  We divide the 
Tully galaxies into the same distance bins, resulting in five non-zero
samples. The number 
of the included galaxies, the distance range of each Tully sample are 
listed in the 1st and 2nd columns of Table 1.   For each Tully sample,  
we carry out the uniform redistribution of galaxy inclination angles 
within each Hubble type separately.

Now that we have independently measured tidal shear and galaxy spin fields, 
we can estimate the strength of intrinsic correlations between the tidal 
shear and galaxy spin fields. 

\section{MEASUREMENTS OF INTRINSIC SHEAR EFFECTS}

Lee \& Pen (2000, 2001) have shown that the intrinsic shear effect on
galaxy spins can be quantified  by one dimensionless parameter, $a$
in range of $[0,0.6]$, and derived its optimal estimation formula: $a =
2 - 6\sum_{i=1}^{3}\hlam_{i}^{2}\hL_{i}^{2}$ where $\{\hlam_{i}\}$ are
the three eigenvalues of shear, and $(\hL_i)$ is the unit spin vector
measured in the shear principal axis frame.  The associated statistical
uncertainty, $\sigma$, can be calculated as the standard deviation of $a$
assuming the null hypothesis that the orientations of galaxy spins are
completely random with no intrinsic alignment: $\sigma = 2/\sqrt{5N}$
where $N$ is the total number of independent unit spin vectors \cite{lp01}.
Each galaxy is used twice, once for each degenerate orientation.
We have verified numerically that using $N$ given by twice the number of
galaxies correctedly reproduces the standard deviation for an uncorrelated
sample.

We calculate the values of $a$ at the positions of all Tully galaxies
using this formula, and averaged them within each sample separately.
Table 1 lists the observed signal $a^{obs}$ of each sample in the 3rd
column. We have also checked the effect of measurement errors in the 
axis ratios and position angles of Tully galaxies on the observed signal 
of intrinsic shear effect. Given the average mean errors of axis ratio 
and position angle \cite{vau-etal91}, we have calculated the propagation 
error due to measurement inaccuracy and found it to be small, 
less than $10\%$ of the signal for every sample. 
Fig. 1 illustrates a  statistical sample of intrinsic shear effect 
on the Tully unit spins in the distance range $[98-100]$ $h^{-1}$ Mpc. 

The observed signal does not give the full strength of intrinsic 
shear effect. In our analysis the shear field has been convolved by 
the adaptive Wiener-filter to minimize noise, but the full strength of 
intrinsic shear effect should be measured from the shear field smoothed 
with a top-hat galactic radius.  
The observed signal, $a^{obs}$, can be rescaled to the full signal, $a$, 
by $a = a^{obs}[{\rm cov}(\del_{w},\del_{th})]^{-2}$. 
Here ${\rm cov}(\del_{w},\del_{th})$ is the covariance between wiener 
filtered and top-hat filtered density fields ($\del_{w}$,$\del_{th}$ 
respectively). 
Approximating the density power spectrum on galactic scales as a power 
law $P(k) = k^{-2}$, and using a top-hat radius of $0.55$$h^{-1}$ Mpc 
which corresponds to the typical size of a galaxy halo \cite{bar-etal86}, 
we rescale the observed signals from each sample separately and find 
the full signal of intrinsic shear effect, which are listed in the 4th column 
of Table 1.  This results from each Tully sample except for the 1st one are
consistent with one another, detecting a nonzero signal of a few
percent.  Therefore, the null hypothesis that the orientations of 
galaxy spin axes in the local shear frame are completely random 
with no intrinsic alignment  can be safely abandoned.  

In the null hypothesis, the distribution of the correlation 
parameter for each sample is Gaussian with zero mean. The non-zero 
value of $a_i$ ($i=1,\cdots,5$) of the Tully sample would just be 
statistical fluctuations with standard deviation $\sig_i$.  
Treating each sample as independent, we define five independent 
variables, $A_1,\cdots, A_5$  as $A_i \equiv \int_{a_{i}}^{\infty}da \frac{1}
{\sqrt{2\pi}\sig_{i}}{\rm exp}(-\frac{a^2}{2\sig_{i}^2})$. 
In the null hypothesis 
each $A_i$ is a random variable, uniformly distributed in the 
range $[0,1]$.  The probability of the null hypothesis, that is, the 
probability that the Tully sample doe not detect intrinsic alignments of galaxy 
spins is given as the cumulative probability, $P(< W)$ with 
$W \equiv \Pi_{i=1}^{5}A_i$.   One can show that this probability is given 
as $P(W) = [{\rm log}(W)]^{4}/4!$.   From the results of Table 1, we find 
$W = 3.25\times 10^{-8}$ and $P(< W) = 1.53\times 10^{-4}$, which rules 
out the null hypothesis at the confidence level higher than $99.98\%$. 

Finally we compute the weighted average of the signals by $\bar{a}
= \frac{\sum_{i}(a_{i}/\sig_{i}^{2})}{\sum_{i}(1/\sig_{i}^{2})}$
and $\bar{\sig} = [\sum_{i}(1/\sig_{i}^{2})]^{-1/2}$.  Excluding the
inconsistent result from the 1st sample, and we find $\bar{a} = 0.17
\pm 0.04$, which is a 25\% accurate measurement of the intrinsic shear
alignment effect. 

\section{DISCUSSIONS AND CONCLUSIONS}

We have presented an observational evidence for the existence of intrinsic
shear effect on present-day galaxy spin orientations.  We have used
current observational data (the PSCz and the Tully galaxy catalogs)
to measure the tidal shear and the galaxy spin fields independently,
and measured the strength of intrinsic shear effect directly.  We have
used five samples in different distant range and measured the signal
of each sample separately.  All the samples except for the innermost
one have produced non-negligible signal of intrinsic shear effects on
galaxy spins, and thus we have been able to rule out the null hypothesis
that galaxy spins are randomly aligned in the tidal shear field at a
confidence level higher than $99.98\%$.  The mean strength of the signal
averaged over the four samples except for the innermost one has been
shown to be $\bar{a}=0.17\pm 0.04$.  The reconstruction procedure is likely 
to systematically underestimate $\bar{a}$.  

Yet, there are two statistical problems in the measurement of 
$\bar{a}$, which we would like to address before concluding.  
The first concern is about excluding the 1st sample.  
When combining the samples that have high signal-to-noise ratio, 
we should give each sample a weight proportional to its volume.  
In such a weighting, the 1st Tully sample has the smallest weight.  
Furthermore, structures are coherent on scales of many Mpc, so that the 
``cosmic variance''is expected to be large in the small sample.  
Our estimator for $a$ assumed that the shear is known without any errors, 
i.e. we neglected``cosmic variance''.   This first sample has a small 
statistical error, but it is not clear how to include this sample when we 
in fact know that the shear field is not known to nearly such good accuracy.  
Cosmic shear has a correlation length comparable to the size of this bin, 
so the coherent errors in this bin are expected to be large.  Hence, 
given the potentially large contributions of cosmic variance, 
we exclude it in the calculation of $\bar{a}$.  

Excluding the 1st sample, 
however, is the simplest way to treat the effects of cosmic variance on 
$\bar{a}$.  For a more realistic treatment,  one should include the 
cosmic variance error, assuming a particular relation between the sample 
size and the value of $a$, and use a maximum likelihood analysis for 
$\bar{a}$. The second concern is about the assumed statistical model for 
$\bar{a}$. Here we have assumed a null hypothesis of $\bar{a}$ equal zero, 
and have shown that alignments are not random at the $99.98\%$ level.  
If cosmic variance is indeed the origin for the small apparent signal in 
the first bin, the estimate of $\bar{a}$ has an additional uncertainty from
cosmic variance.  The quantitative estimate of the cosmic scatter,  
however, is beyond the scope of this letter.
 
We finally conclude that the null hypothesis of no intrinsic alignment 
is very safely ruled out, and the resulting signal represents the average 
strength of intrinsic shear effect from the galaxy samples that have 
consistent nonzero signals.  Our result is consistent with the linear 
tidal torque model based on the gravitational instability. 
It also hints that the present galaxy spin field might be used as a probe 
to measure the initial tidal shear and mass density fields, and eventually 
map the three dimensional matter distribution of the universe \citep{lp01}.  
Applying the results of \citep{cri-etal01}, we conclude that weak lensing 
surveys are strongly contaminated by this effect at low redshift, but only 
negligibly at high redshift.

\acknowledgments
We thank B. Tully and E. Branchini for their catalogs and useful comments. 
We also thank the referee for helping us improve the manuscript. 
This work has been supported by the Academia Sinica and NSERC.

\newpage
\begin{table}
\caption{Observed and rescaled signal of intrinsic shear}
\begin{center}
\begin{tabular}{cccc}\hline \hline
\# of galaxies & $r$ & $a^{obs}$ & $a$ \\
& ($h^{-1}$Mpc) & \\
\hline   
5727& 0-50 & $(0.1 \pm 0.8)\times 10^{-2}$  & $0.003 \pm 0.03$\\ 
3336& 50-75 & $(3.2 \pm 1.1)\times 10^{-2}$ & $0.14 \pm 0.05$\\
1940& 75-100 & $(4.6 \pm 1.4)\times 10^{-2}$ & $0.24 \pm 0.07$\\
898& 100-125 & $(1.4 \pm 2.1)\times 10^{-2}$ & $0.10 \pm 0.14$\\
221& 125-150 & $(3.0 \pm 4.3)\times 10^{-2}$ & $0.25 \pm 0.34$\\ 
\hline 
\end{tabular}
\end{center}
{\em Note:\/} $\bar{a}$ has been found to be $0.17 \pm 0.04$. 
The signal of the first sample was not used in the calculation of $\bar{a}$, 
since its statistical weight should be dominated by cosmic 
variance.
\end{table}

\clearpage
\begin{figure}
\plotone{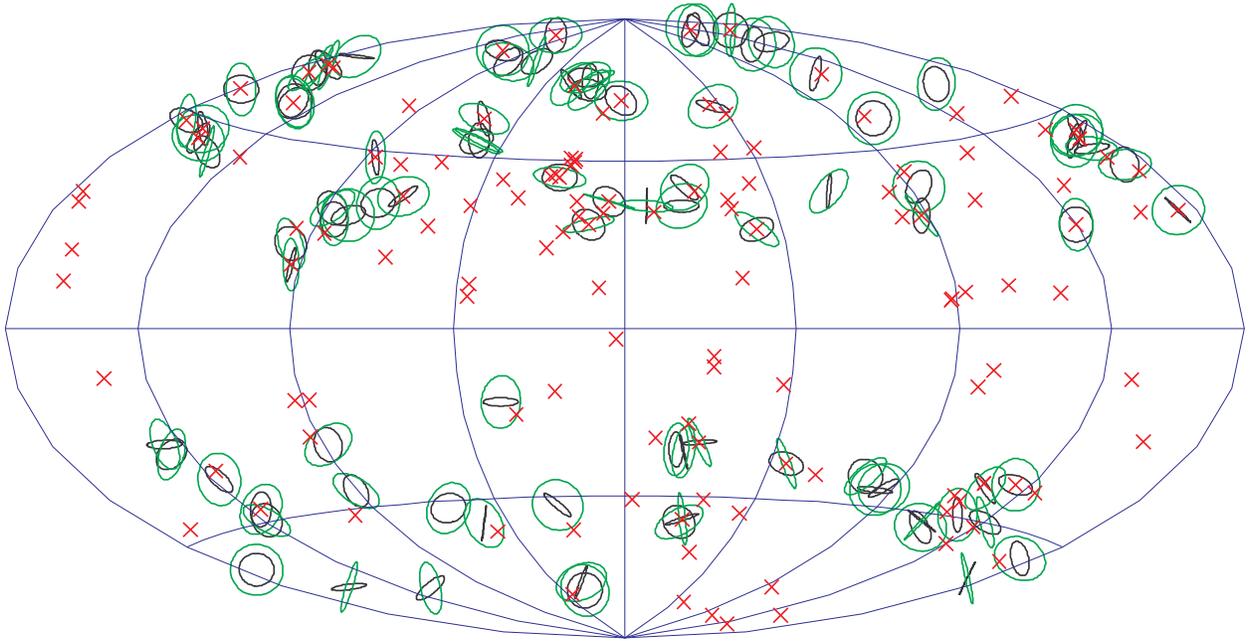}
\caption{The two-dimensional illustration for different realizations of 
intrinsic shear effect.  It is an Aitoff projected sky map in galactic 
coordinates within distance range [98-100] $h^{-1}$ Mpc. 
The smaller black ellipses correspond to the actual alignment of disk 
galaxies while the green larger ellipses represent the expectation value 
of predicted disk alignments where spins are perfectly aligned with  the 
intermediate principal axes of tidal shears. 
The black ellipses are determined using the observed position angles 
and axis ratios rescaled to the thin disk model. The red crosses show 
the positions of PSCz galaxies in the same distance range.}
\end{figure}
\end{document}